\documentclass[12pt]{article}
\usepackage{amsmath}
\usepackage{amssymb}
\usepackage{epsfig}
\usepackage{cite}

\newcommand\npb[3]{{\it Nucl.\ Phys.\ }{\bf B #1} (#2) #3}
\newcommand\plb[3]{{\it Phys.\ Lett.\ }{\bf B #1} (#2) #3}
\newcommand\jhep[3]{{\it J. High Energy Phys.\ }{\bf #1} (#2) #3}

\newcommand\pre[3]{{\it Phys.\ Rev.\ }{\bf E #1} (#2) #3}

\newcounter{hran}
\renewcommand{\thehran}{\arabic{hran}}

\def\bminiG#1{\setcounter{hran}{\value{equation}}
\refstepcounter{hran}\setcounter{equation}{-1}
\renewcommand{\theequation}{\thehran\alph{equation}}
\refstepcounter{equation}\label{#1}\begin{eqnarray}}

\def\emini{\end{eqnarray}\relax\setcounter{equation}{\value{hran}}\renewcommand{\theequation}{\arabic{equation}}}

\def\as{\alpha_{\mbox{\scriptsize s}}}

\def\Re{\mathop{\rm Re}}

\def\0{{\rm\bf 0}}
\def\sing{{\rm\bf 1}}
\def\s{{\rm\bf s}}
\def\a{{\rm\bf a}}
\def\8{{\rm\bf 8}}
\def\10{{\rm\bf 10}}
\def\27{{\rm\bf 27}}

\def\al{\alpha}
\def\be{\beta}
\def\Gam{\Gamma}

\def\cP{{\cal P}}
\def\cQ{{\cal Q}}
\def\cS{{\cal S}}
\def\cV{{\cal V}}

\def\cO#1{{\cal{O}}\left(#1\right)}
\def\half{\mbox{\small $\frac{1}{2}$}}

\def\abs#1{\left|#1\right|}
\def\Tr{{\rm Tr}}

\begin{document}

\begin{flushright}
  Bicocca--FT--05--20\\
     LPTHE-05-21  \\
     hep-ph/0508130\\
     August 2005
\end{flushright}

\begin{center}
{\Large\bf Hadron collisions and the fifth form factor}

\vspace{7mm}

{\large Yu.L.\ Dokshitzer\footnote{On leave from St.\ Petersburg
Nuclear Institute, Gatchina, St.\ Petersburg 188350, Russia}$^{,\rm
a}$ and G.\ Marchesini$^{\rm a, b}$}
\end{center}

\vspace{1mm}

 {\small a)  LPTHE, Universit\'es Paris-VI-VII, CNRS UMR
7589, Paris, France  }

{\small b) Dipartimento di
Fisica, Universit\`a Milano-Bicocca and INFN, Sezione Milano, Italy}

\vspace{8mm}

\centerline{\small\bf Abstract}
\begin{quote}
  Logarithmically enhanced effects due to radiation of soft gluons at
  large angles in $2\to 2$ QCD scattering processes are treated in
  terms of the "fifth form factor" that accompanies the four collinear
  singular Sudakov form factors attached to incoming and outgoing hard
  partons. Unexpected symmetry under exchange of internal and external
  variables of the problem is pointed out for the anomalous dimension
  that governs soft gluon effects in hard gluon--gluon scattering.
\end{quote}

\section{Cross-channel colour transfer and soft gluons}

Measuring final--state characteristics in hard hadron--hadron
interactions supplements the overall hardness scale $Q$ of the
underlying parton scattering process $p_1,p_2\to p_3 ,p_4$ with the
second (hard) scale $Q_0\ll Q$ that quantifies small deviation of the
final state system from the Born kinematics (out-of-event-plane
particle production, near-to-backward particle correlations, inter-jet
energy flows, etc.).  The ratio of these two scales being a large
parameter calls for analysis and resummation of double (DL) and single
logarithmic (SL) radiative corrections in all orders.  Logarithmically
enhanced (both DL and SL) contributions of {\em collinear}\/ origin
are easy to analyse and resum into exponential Sudakov form factors
belonging to hard primary partons.

SL effects due to soft gluons radiated at large angles pose more
problems. For example, particle energy flow $E=Q_0$ in a given
inter-jet direction acquires contributions $\cO{\as^n\ln^n(Q/Q_0)}$
from ensembles of $n$ energy ordered gluons radiated at arbitrary
(large) angles. In general, such ``hedgehog'' multi-gluon
configurations contribute at the SL level to the so called {\em
non-global}\/ observables that acquire contributions from a restricted
phase space window ~\cite{ng} and are difficult to analyse.
On the contrary, {\em global observables}\/ that acquire contributions
from the full phase space are free from this trouble: only the hardest
among the secondary gluons contributes while the softer ones don't
affect the observable so that their contributions cancel against
corresponding virtual terms.  As a result,
the problem reduces to the analysis of {\em virtual corrections}\/ due
to multiple gluons with $k_t>Q_0$ attached to primary hard partons
only.  They can be treated iteratively and fully exponentiated,
together with DL terms.

The programme of resumming soft SL effects due to large angle gluon
emission in hadron--hadron collisions was pioneered by Botts and
Sterman~\cite{BS} (see also \cite{KOS,BCMN,ASZ}).  Complication arises
from the fact that gluon emission changes the {\em colour state}\/ of
the hard parton system which in turn affects successive radiation of a
softer gluon.

In this letter we propose a general treatment of large angle gluon
radiation. It is based on the observation that the square of the
eikonal current for emission of gluon $k=(\omega,{\bf k})$ off an
ensemble of four partons $i=1,\ldots 4$,
\begin{eqnarray}
\label{eq:jdef}
   j^{\mu,b}(k) = \sum_{i=1}^{4} \frac{\omega \, p_i^\mu}{(k
     p_i)}T_i^b\,; \qquad
    \sum_{i=1}^{4} T_i^b \>=\> 0,
\end{eqnarray}
can be represented as
\begin{equation}
\label{eq:alter}
\begin{split}
  - j^2(k) =\> & T_1^2\, W^{(1)}_{34}(k) + T_2^2\, W^{(2)}_{34}(k) +
  T_3^2\, W^{(3)}_{12}(k) +  T_4^2\, W^{(4)}_{12}(k)  \\
  &+\> T_t^2\cdot A_{t}(k) \> \>+ T_u^2\cdot A_u(k)\,.
\end{split}
\end{equation} 
Here $T_i^2$ is the $SU(N)$ ``colour charge'' of parton $p_i$ and the
two operators $T_t^2$ and $T_u^2$ are the charges exchanged in the
$t$- and $u$- channels of the scattering process,
\begin{equation}
\label{eq:TT}
 T_t^2 = (T_3+T_1)^2 = (T_2+T_4)^2, \quad T_u^2= (T_4+T_1)^2= (T_2+T_3)^2.
\end{equation} 
The functions $W$ are combinations of {\em  dipole antennae}\/
\begin{equation}\label{eq:W123def}
   W^{(1)}_{34} = w_{13}+w_{14}-w_{34}\,, 
   \quad w_{ij}(k)=  \frac{\omega^2\,(p_ip_j)}{(kp_i)(kp_j)} . 
\end{equation}
The distribution \eqref{eq:W123def} is collinear singular {\em only}\/
when ${\bf k} \parallel {\bf p}_1$. This singularity contributes
proportional to the corresponding Casimir operator, in accord with
general factorisation property.  Integrating \eqref{eq:W123def} over
angles gives 
\begin{equation}\label{eq:Wangint}
\int\frac{d\Omega}{4\pi}\> W ^{(1)}_{34} \>=\>
   \ln\frac{(p_1p_3)(p_1p_4)}{(p_3p_4)\, m^2}  = \ln\frac{tu}{2m^2\,s}\,.
\end{equation}
with $m^2$ the collinear cutoff and $s=(p_1\!+\!p_2)^2,\>
t=(p_1\!-\!p_3)^2,\> u=(p_1\!-\!p_4)^2$.  The collinear cutoff $m$
disappears when the virtual and real contributions (to a collinear and
infrared safe observable) are taken together, and gets replaced by the
proper observable dependent scale ${Q_0}$, see examples
in~\cite{QVscale}.

The first four terms in \eqref{eq:alter} are collinear singular and
belong to individual partons; their colour factors are numbers.  Their
exponentiation leads to the product of four DL Sudakov form factors
$F_i(Q_0,Q)$ with the common hard scale $Q^2= {tu}/{s}=
s\sin^2\Theta_s$.

The angular dipole combinations in the last two terms in
\eqref{eq:alter} are
\begin{equation}
\label{eq:AtAu}
A_t= w_{12} +w_{34}-w_{13}-w_{24}\,,\quad 
A_u= w_{12} +w_{34}-w_{14}-w_{23}\,.
\end{equation} 
Unlike the dipoles $W^{(i)}_{jk}$, they are integrable in angles:
\begin{equation}
\label{eq:AtAuangint}
\int\frac{d\Omega}{4\pi}\, A_t(k) \>=\>  2\ln\frac{s}{-t}\,; \qquad 
\int\frac{d\Omega}{4\pi}\, A_u(k) \>=\>  2\ln\frac{s}{-u}\,.
\end{equation} 
Contrary to the first four DL contributions,  
this additional contribution originates from coherent gluon radiation
at angles {\em larger than the cms scattering angle $\Theta_s$} and
gives rise to the {\em fifth form factor}\/ $F_X(\tau)$ which is a
single logarithmic function of the parameter
\begin{equation}
  \label{eq:tau}
 \tau = \int_{Q_0}^{Q}\frac{dk_t}{k_t}\frac{\as(k_t)}{\pi}\,.
\end{equation}
Virtual corrections to the hard scattering matrix element are
calculated simply by exponentiating a minus half of the real eikonal
emission probability \eqref{eq:alter}. Additional Coulomb corrections
due to virtual gluon exchanges between two incoming or two outgoing
partons are obtained from \eqref{eq:alter} by keeping only $s$-channel
interference terms $w_{12}$ and $w_{34}$ and replacing them by $i\pi$.
The finite non-Abelian Coulomb phase\footnote{a divergent part of the
  Coulomb phase, though non-Abelian, factors out and cancels in
  observables~\cite{DMunder}} can be simply incorporated by adding the
phases to the logarithms in \eqref{eq:AtAuangint}.

In summary, virtual dressing of the hard matrix element results in 
\begin{equation}
  \label{eq:dress} 
  M_0 \>\Longrightarrow\> \prod_{i=1}^4 F_i(Q_0,Q) \cdot M(\tau), \qquad 
  M(\tau) = F_X(\tau)\cdot M_0\,,
\end{equation}
where $F_X$ (with subscript $X$ standing for ``cross-channel'') is given by
\begin{equation}
  \label{eq:Fx} F_X(\tau) \>=\> \exp\left\{ -\tau \left( T^2_t\cdot
  T + T^2_u\cdot U\right)\right\}, \quad T=\ln\frac{s}{t}=
  \ln\frac{s}{-t} -i\pi, \>\> U=\ln\frac{s}{u}\,.
\end{equation}
This expression for the fifth form factor holds for scattering of
arbitrary colour objects. 
It is present in the QED context as well (where it is determined by
electric charge transfers in cross channels), however in QCD the
operators $T_t^2$ and $T_u^2$ do not commute and this is what
complicates the analysis.

The two scale distributions acquire the {\em soft factor}\/ $\cS_X$:
\begin{equation}
\label{eq:Sig}
   \Sigma(Q_0,Q) \>=\> \Sigma^{{\mbox{\scriptsize coll}}}(Q_0,Q) \cdot
   \cS_X(\tau), \qquad \cS_X(\tau) \>=\> \frac{\Tr
   (M^\dagger(\tau)\>M(\tau))}{\Tr (M_0^\dagger\> M_0)}\> \cong
   \abs{F_X}^2 .
\end{equation}
Here $\Sigma^{{\mbox{\scriptsize coll}}}$ embodies the first four
terms in \eqref{eq:alter} as well as collinear logarithms from parton
distribution functions.

It is convenient to work in the colour basis of $s$-channel projectors
$\cP_\al$ onto the irreducible $SU(N)$ representations $\al$ present
in the colour space of the two incoming partons ($p_1,p_2$).  The
colour transfer matrices $T_t^2$ and $T_u^2$ in \eqref{eq:Fx} can be
easily found with use of the re-projection matrices\footnote{For
  $gg\to gg$ these matrices are given in Appendix, see
  \eqref{App:Kts}.}  that express $t$- and $u$-channel colour
projectors in terms of $s$-channel ones, $ \cP^{(t)}=K_{ts}\cdot \cP$,
$ \cP^{(u)}=K_{us}\cdot \cP$, and vice versa, $K_{st}=K_{ts}^{-1}$,
$K_{su}=K_{us}^{-1}$.  We have
\begin{equation}
\label{eq:key}
  T_t^2 \>=\> K_{st}\, C_2^{(t)}\, K_{ts}, \quad T_u^2 \>=\> K_{su}\,
  C_2^{(u)}\, K_{us},
\end{equation}  
with $C_2$ the diagonal matrix of Casimir operators of all irreducible
representations present in the $t$ ($u$) channel.  An advantage of the
representation of the {\em soft anomalous dimension}\/
\begin{equation}
\label{eq:andim}
   \Gamma \>=\> -\left(T^2_t\cdot T + T^2_u\cdot U\right) \>\> \equiv
   -N(T+U)\cdot \cQ
\end{equation}
in terms of cross-channel charges \eqref{eq:key} is trivialisation of
the analysis of the Regge behaviour. In the case of small angle
scattering one term dominates, $T\gg U$ (forward scattering) or $U\gg
T$ (backward), and the anomalous dimension $\Gam$ becomes diagonal in
the corresponding channel so that the problem becomes essentially
Abelian. Resulting exponents are nothing but Regge trajectories of
$t$-($u$-) channel exchanges that are proportional to corresponding
Casimirs.

\section{Gluon--gluon scattering}

As an example we considered in detail the case of gluon--gluon
scattering that was first treated by Kidonakis, Oderda and 
Sterman in~\cite{KOS} and turned out to be rather complicated 
since it  involves many colour  channels (six for $SU(N)$).

Characterised in terms of  irreducible representations, 
two gluons in $SU(3)$ can be in the states
\begin{equation}
  \label{eq:irred3}
  \bf{glue} \otimes \bf{glue} 
  = \bf{8}_\a + \10 + \sing + \bf{8}_\s + \27,
\end{equation}
where $\bf{8}_a$ and $\10$ mark {\em antisymmetric}\/ representations
(with corresponding dimensions) and three {\em symmetric}\/ ones are
the singlet ($\sing$), octet ($\bf{8}_s$) and the high symmetric
tensor representation ($\27$).  In the general case of $SU(N)$ (with
$N>3$) we have an additional {\em symmetric}\/ representation (which
we mark \0):
\begin{equation}
  \label{eq:irredN}
  \bf{glue} \otimes \bf{glue} 
  = \bf{8}_\a + \10 + \sing + \bf{8}_\s + \27 +\0.
\end{equation}
We use the $SU(3)$ motivated names in spite of the
fact that the dimensions of corresponding representations are actually
different from 8, 10, etc.:
\begin{equation}
\label{eq:Fs}
\begin{split}  
  K_\sing &= 1, \qquad  K_\a     = K_\s = N^2-1, \qquad
  K_{\10} = 2\cdot \frac{(N^2-1)(N^2-4)}{4}, \\
  K_{\27} &= \frac{N^2(N-1)(N+3)}{4}, \qquad\qquad \quad
  K_{\0}  \>=  \frac{N^2(N+1)(N-3)}{4}.
\end{split}
\end{equation}
The $s$-channel projectors $\cP_\al$ are explicitly constructed in
\cite{DMunder}. The projector basis we order as follows:
\begin{equation} \label{eq:basis}
  \cP_\al=\{\cP_\a,\, \cP_\10\,, \cP_\sing,\, \cP_\s,\, \cP_\27,\,
  \cP_\0\}\,.
\end{equation}
The corresponding Casimir operators read
\begin{equation}
\label{eq:C2matr}
(T^a)^2_{\al\be} \>=\> (C_2)_{\al\be} = \delta_{\al\be} \cdot c_\al
\,, \qquad c_\al=
\{N,\>2N,\>0,\>N,\>2(N\!+\!1), \>2(N\!-\!1)\}.
\end{equation}
where $C_2$ is the diagonal matrix entering \eqref{eq:key}.

The normalised anomalous dimension $\cQ$ defined in \eqref{eq:andim}
is a $6\times 6$ matrix that depends on $N$ and the ratio of
logarithms $T/U$ as unique geometry dependent (complex) parameter.  It
has six eigenstates. Three simple, $N$-independent, eigenvalues are 
\begin{equation}
  \label{eq:E123} 
  E_1 = 1\,, \quad E_2 = \frac{3-b}2\,, \quad E_3 = \frac{3+b}2\,;
  \qquad b\>\equiv\>\frac{T-U}{T+U}\,.
\end{equation}
Three $N$-dependent eigenvalues satisfy the cubic equation 
\begin{equation}
\label{eq:E13form}
\left[E_i-\!\frac{4}{3}\right]^3 
- \frac{(1+3b^2)(1+3x^2)}{3}\left[E_i-\!\frac43\right] 
- \frac{2(1-9b^2)(1-9x^2)}{27}  \>=\> 0; \qquad  x\>\equiv\> \frac1N\,.
\end{equation}
Its solutions can be parametrised as follows:
\bminiG{456}
  \label{eq:E456}
  E_{4,5,6} &=& \frac43\left(1 \,+\, \frac{\sqrt{(1+3b^2)(1+3x^2)}}{2}\>
  \cos\left[\frac{\phi + 2k\pi}{3} \right]\right), \quad k = 0,\,1,\,
2\,; \qquad { } \\
\label{eq:R}
 \cos\phi&=&R\,, \qquad 
 R = \frac{(1-9b^2)(1-9x^2)}
{\left[(1+3b^2)(1+3x^2)\right]^{3/2}}\,.
\emini
In our representation the three $N$-dependent energy levels \eqref{456} 
and corresponding eigenvectors 
are explicitly real functions of $b$ (the property not easy to
extract from \cite{KOS}). 
We give explicit expression for the soft factor $\cS_X$ in special
cases when eigenvalues \eqref{456} simplify.

\subsection{Scattering at $90^{\rm o}$}

The simplest case is $90$ degree scattering which corresponds to
$b\!=\!0$ ($t=u$).  Here $\cQ$ \eqref{eq:Qdef} is diagonal so that the
$s$-channel projectors $\cP_{\al}$ become eigenvectors whose
eigenvalues are just the corresponding diagonal elements of $\cQ$:
\begin{eqnarray}
  \label{eq:E123456-b=0}
E_k &=& \left\{ 1,\> \frac32,\> \frac32,\> 2,\> \frac{N\!-\!1}{N},\>
\frac{N\!+\!1}{N}
\right\} ,\\ 
\cV_\kappa &\propto& 
 \left\{ \cP_\10,\> \cP_\s+\cP_a,\> \cP_\s-\cP_a,\> \cP_\sing,\>
   \cP_\27,\> \cP_\0 \right\} .
\end{eqnarray}
To present the answer for the soft factor $\cS_X$ we define the
suppression factors
\begin{equation}
  \chi_t(\tau) \>=\> \exp\left\{-2N\tau \cdot\ln\frac{s}{-t}\right\},
  \quad \chi_u(\tau) \>=\> \exp\left\{-2N\tau
  \cdot\ln\frac{s}{-u}\right\}.
\end{equation}
In $90^o$ scattering kinematics we have $\chi_t=\chi_u$, $\Re T=\Re U=
\ln 2$ and we get
\begin{equation} \label{eq:Sb0}
 \cS_X(\tau) = \frac{\chi^2}{3}\left[\, \frac{4\, \chi^{2}}{N^2\!-\!1} +
 { \chi} + \frac{N\!-\!3}{N\!-\!1}\, \chi^{\frac2N} +
 \frac{N\!+\!3}{N\!+\!1}\,
\chi^{-\frac2N} \,\right], \qquad \chi(\tau)\>=\>
\exp\left\{-2N\tau\ln2\right\}.
\end{equation}

\subsection{$N\to\infty$ limit}

Now the eigenvalues of $\cQ$ are as follows: 
$$
 E_1= 1, \quad  E_2= \frac{3-b}{2}, \quad  E_3= \frac{3+b}{2}, 
 \quad  E_4= 2, \quad  E_5= 1-b, \quad  E_6= 1+b\,,
$$
and the corresponding eigenvectors are 
\begin{equation}
\label{eq:Vx0}
\cV_{1, \ldots 6} \>=\>   
\left[ \begin {array}{r} 0\\\noalign{\medskip}0\\\noalign{\smallskip}0
\\\noalign{\medskip}0\\\noalign{\medskip}1\\\noalign{\smallskip}-1
\end {array} \right] 
\left[ \begin {array}{r} {1+b}\\\noalign{\medskip} -2b
\\\noalign{\medskip}0\\\noalign{\medskip}  {1+b}
\\\noalign{\medskip}-b\\\noalign{\medskip}-b \end {array} \right] 
\left[ \begin {array}{r} {1-b}\\\noalign{\medskip} 2b
\\\noalign{\medskip}0\\\noalign{\medskip} {-1+b}
\\\noalign{\medskip}-b\\\noalign{\medskip}-b\end {array} \right] 
 \left[ \begin {array}{r} -4(1\!-\!{b}^{2}) \\\noalign{\medskip}-8
 {b^3}
\\\noalign{\medskip} (1\!-\!{b}^{2})^{2}
\\\noalign{\medskip} 4 (1\!-\!{b}^{2})
\\\noalign{\medskip} 2b^2(1\!+\!b^2) \\\noalign{\medskip} 2b^2(1\!+\!b^2)
\end {array} \right] 
 \left[ \begin {array}{r} 0\\\noalign{\medskip}2\\\noalign{\medskip}0
\\\noalign{\medskip}0\\\noalign{\medskip}1\\\noalign{\medskip}1
\end {array} \right] 
 \left[ \begin {array}{r} 0\\\noalign{\medskip}-2\\\noalign{\medskip}0
\\\noalign{\medskip}0\\\noalign{\medskip}1\\\noalign{\medskip}1
\end {array} \right]. 
\end{equation}
The soft factor becomes
\begin{equation}
\label{eq:x0ans}
  \cS_X(\tau) \>=\> {\chi_t \, \chi_u}\> \frac{ (m_t+m_u)^2 +
  (m_s-m_u)^2\,\chi_t + (m_s+m_t)^2\,\chi_u}{(m_t+m_u)^2 + (m_s-m_u)^2
  + (m_s+m_t)^2}.
\end{equation}
Here $m_s$, $m_t$, $m_u$ are pieces of the Born $gg$ scattering matrix
element each containing the gluon exchange diagram in the
corresponding channel (together with the piece of the four-gluon
vertex with the same colour structure). The combinations $m_t+m_u$,
$m_s+m_t$ and $m_s-m_u$ are gauge invariant amplitudes,
$$
 (m_t+m_u)^2 \>=\>  1-\frac{st}{u^2}-\frac{us}{t^2} +\frac{s^2}{tu} ,
$$
with the other two obtained by crossing symmetry. 
The squared matrix element reads
$$ \Tr( M_0^\dagger\> M_0) \>=\> \half{ N^2(N^2-1)} \left[\,
(m_t+m_u)^2 + (m_s-m_u)^2 + (m_s+m_t)^2 \right].
$$

\subsection{Regge limit}

In the case of close to forward scattering, $|t|\ll s\simeq |u|$, we
have $b\to1$ at which point eigenvalues \eqref{eq:E456} also
simplify. Since $U$ in \eqref{eq:andim} is negligible, according to
\eqref{eq:key} the eigenvectors coincide with $t$-channel projectors,
\begin{equation}
\begin{split}
  \cV_1 & =K_{st}\cdot\cP^{(t)}_\a,\qquad
  \cV_2=K_{st}\cdot\cP^{(t)}_\s, \qquad
  \cV_3  =K_{st}\cdot\cP^{(t)}_\10,\\
  \cV_4 & =K_{st}\cdot\cP^{(t)}_\sing, \qquad
  \cV_5  =K_{st}\cdot\cP^{(t)}_\0,\qquad \cV_6\,.
  =K_{st}\cdot\cP^{(t)}_\27,
\end{split}
\end{equation}
and the eigenvalues with the corresponding Casimirs \eqref{eq:C2matr}:
\begin{equation}
 \{E_\kappa\} \>=\> \frac{1}{N}\left\{ c_\a,\, c_\s,\, c_\10,\,
 c_\sing,\, c_\0,\, c_\27 \right\}.
\end{equation}
Since for $t\to 0$ scattering in the Born approximation is dominated
by $t$-channel one gluon exchange, we are left with 
\begin{equation}
\label{eq:b1ans}
    S(\tau)\>=\> \chi_t(\tau) \>=\> \left(\frac{s}{t}\right)^{-2N\tau},
\end{equation}
which exponent coincides with the (twice) Regge trajectory of the
gluon.

\section{Final remark}

The cubic equation \eqref{eq:E13form} for the $N$-dependent energy
levels 4, 5, 6 of $\cQ$ possesses a weird symmetry which interchanges
internal (colour group) and external (scattering angle) degrees of
freedom:
\begin{equation} \label{eq:weird2}
  \frac{T+U}{T-U} \quad \Longleftrightarrow \quad  N\,.
\end{equation} 
In particular, this symmetry relates $90$-degree scattering, $T=U$,
with the large-$N$ limit of the theory.
Giving the complexity of the expressions involved, such a symmetry being
accidental looks highly improbable. Its origin remains mysterious and
may point at existence of an enveloping theoretical context that
correlates  internal and external variables (string theory?).

\appendix

\section{technicalities}

\paragraph{Soft anomalous dimension.}
The normalised anomalous dimension matrix $\cQ$ has a block structure
and reads
\begin{eqnarray}
\label{eq:Qdef}
 \cQ &=& \left( \begin{array}{rrrrrr} \frac32 & 0 & \>\> -2b & \>\>
 -\frac{1}{2} b & -\frac{2}{N^2}b & -\frac{2}{N^2}b \\[2mm] 0 & 1 & 0
 & -b & \>\> -\frac{(N+1)(N-2)}{N^2}b & \>\> -\frac{(N-1)(N+2)}{N^2}b
 \\[2mm] -\frac{2}{N^2-1}b & 0 & 2 & 0 & 0 & 0 \\[2mm] -\frac12 b&
 -\frac{2}{N^2-4}b & 0 & \frac32 & 0 & 0 \\[2mm] -\frac{N+3}{2(N+1)}b
 & \>\> -\frac{N+3}{2(N+2)}b & 0 & 0 & \frac{N-1}{N} &0\\[2mm]
 -\frac{N-3}{2(N-1)}b & -\frac{N-3}{2(N-2)}b & 0 & 0 & 0 &
 \frac{N+1}{N} \end{array} \right) \qquad { }
\end{eqnarray}
The matrix elements of the states \27 and
\0 (two last rows and columns) are related by the formal operation
$N\to -N$.
Let us remark that our anomalous dimension differs from the one
introduced in~\cite{KOS} by a piece proportional to the unit
matrix. In our approach, this piece is absorbed into the collinear
factor in \eqref{eq:Sig} and determines the common hard scale of the
Sudakov form factors in \eqref{eq:dress} as $Q^2=tu/s$.

\paragraph{Colour re-projection matrices for $gg\to gg$.}
Graphic colour projector technique described in \cite{DMunder} allows
one to find the re-projection matrices $K_{ts}$ introduced in
\eqref{eq:key} without much effort
The matrix $K_{ts}$ is given by
\begin{equation}\label{App:Kts}
K_{ts} = \left( \begin{array}{rrrrrr}
 \frac12  & 0 & 1 & \frac12 & -\frac1N & \frac1N \\[1mm]
 0 & \frac12 & \frac{N^2-4}{2} & -1 & -\frac{N-2}{2N} & -\frac{N+2}{2N} \\[1mm]
 \frac{1}{N^2-1} & \frac{1}{N^2-1} 
&  \frac{1}{N^2-1} & \frac{1}{N^2-1} &
\frac{1}{N^2-1} & \frac{1}{N^2-1} \\[1mm]
 \frac12 & -\frac2{N^2-4} & 1 & \frac{N^2-12}{2(N^2-4)} &
\frac{1}{N+2} & -\frac{1}{N-2}  \\[1mm]
 -\frac{N(N+3)}{4(N+1)} &  \frac{-N(N+3)}{4(N\!+\!1)(N\!+\!2)} 
& \frac{N^2(N+3)}{4(N+1)} &\frac{N^2(N+3)}{4(N\!+\!1)(N\!+\!2)} & 
\frac{N^2+N+2}{4(N\!+\!1)(N\!+\!2)} & \frac{N+3}{4(N+1)}  \\[1mm]
 \frac{N(N-3)}{4(N-1)} & \frac{-N(N-3)}{4(N\!-\!1)(N\!-\!2)} &
 \frac{N^2(N-3)}{4(N-1)} & \frac{-N^2(N-3)}{4(N\!-\!1)(N\!-\!2)} &  
\frac{N-3}{4(N-1)} & \frac{N^2-N+2}{4(N\!-\!1)(N\!-\!2)} 
\end{array}  \right)
\end{equation}
The inverse matrix coincides with the direct one, $K_{st} \>=\>
K_{ts} $.  To construct the $u$-channel re-projection matrices
$K_{us}$ in \eqref{eq:key} one exploits the symmetry of the
$s$-channel projectors under $t\leftrightarrow u$ transformation. Thus
$K_{us}$ is obtained by changing sign of the first two {\em columns}\/
of $K_{ts}$ and the inverse matrix $K_{su}$ --- by changing sign of
the first two {\em rows}\/ of~$K_{ts}$.

\paragraph{Eigenvectors.}

The eigenvectors $\cV_1,\cV_2,\cV_3$ are
\begin{equation}
\label{eq:V123}
{\cal{V}}_{1,2,3} \!=\!  \left[
\begin{array}{c}
 \displaystyle 0  \\[2ex]
 \displaystyle 1  \\[2ex]
 \displaystyle 0 \\[2ex]
 \displaystyle \frac{4b}{N^2\!-\!4}  \\[2ex]
 \displaystyle -\frac{b\,N(N\!+\!3)}{2(N\!+\!2)} \\[2ex]
 \displaystyle \frac{b\,N(N\!-\!3)}{2(N\!-\!2)}
\end{array}
 \right]
\quad
\left[
\begin{array}{c}
 \displaystyle 1\!+\!b  \\[2ex]
 \displaystyle -2b  \\[2ex]
 \displaystyle \frac{4b}{N^2\!-\!1} \\[2ex]
 \displaystyle 1+ \frac{b\,(N^2\!-\!12)}{N^2\!-\!4}  \\[2ex]
 \displaystyle -\frac{b\,N(N\!+\!3)}{(N\!+\!1)(N\!+\!2)}\\[2ex]
 \displaystyle -\frac{b\,N(N\!-\!3)}{(N\!-\!1)(N\!-\!2)} 
\end{array}
 \right]
\quad
\left[ 
\begin{array}{c}
 \displaystyle -1\!+\!b  \\[2ex]
 \displaystyle -2b  \\[2ex]
 \displaystyle  -\frac{4b}{N^2\!-\!1} \\[2ex]
 \displaystyle  1- \frac{b\,(N^2\!-\!12)}{N^2\!-\!4} \\[2ex]
 \displaystyle \frac{b\,N(N\!+\!3)}{(N\!+\!1)(N\!+\!2)} \\[2ex]
 \displaystyle \frac{b\,N(N\!-\!3)}{(N\!-\!1)(N\!-\!2)}
\end{array}
 \right]. 
\end{equation}
The states 2 and 3 are related by the crossing $t\leftrightarrow u$
transformation: $\cV_3$ is obtained from $\cV_2$ by $b\to -b$ and
changing sign of the antisymmetric projectors (first two rows).

The last three eigenvectors are 
\begin{equation}
  \label{eq:V456}
  {{\cal{V}}_{4,5,6}} =  \left[
\begin{array}{c}
 \displaystyle -\frac {4}{N^2}\,(E_i \!-\! 1) \,b \\ [2ex]
 \displaystyle -\frac {N^{2}\! -\! 4}{N^2}\,(E_i\!-\!2) \,b \\ [2ex]
 \displaystyle \frac1{N^2-1}\left[ \left( E_i-\frac{N\!-\!1}{N}\right)
 \left(E_i-\frac{N\!+\!1}{N}\right) -\frac{N^2 \!-\! 5}{N^2}\,
 b^2\right] \\[2ex] \displaystyle \frac {4}{N^2} \,b^{2}\\ [2ex]
 \displaystyle \frac{N}{N+1}\left[ \frac{N\!+\!2}{2\,N} (E_i\!-\!2)
 \left(E_i-\frac{N\!+\!1}{N}\right) - {2\,b^2}\right] \\ [2ex]
 \displaystyle \frac{N}{N-1}\left[ \frac{N\!-\!2}{2\,N} (E_i\!-\!2)
 \left(E_i-\frac{N\!-\!1}{N}\right) - {2\,b^2}\right]
\end{array}
 \right] ,
\end{equation}
with $E_i$ the corresponding energy eigenvalue, $i=4,5,6$. 

Vectors $\cV_i$ are orthogonal with respect to the scalar product
defined by the metric tensor $W_{\al\be} = K_\al \delta_{\al\be}$,
$$ 
 \left\langle \cV_i \right| W^{-1} \left| \cV_k\right\rangle \>=\>
 0\,, \quad i\neq k\,,
$$
while the matrix $\cQ$ becomes symmetric \cite{DMunder} under
the corresponding metric transformation
$$
W^{-1/2}\,\cQ\,W^{1/2} \>=\> \cQ_{\scriptstyle{sym}} \,.
$$


\begin{thebibliography}{99}

\bibitem{ng} 
M.\ Dasgupta and G.P.\ Salam, 
%``Resummation of non-global QCD observables,''
\plb{512}{2001}{323} [hep-ph/0104277]; \\
 M.\ Dasgupta and G.P.\ Salam, 
%``Accounting for coherence in interjet E(t) flow: A case study,''
\jhep{03}{2002}{017} [hep-ph/0203009]

\bibitem{BS} J.\ Botts and G.\ Sterman, \npb{325}{1989}{62}
  
\bibitem{KOS} N.\ Kidonakis and G.\ Sterman, \plb{387}{1996}{867},
  \npb{505}{1997}{321} [hep-ph/9705234];
  N.\ Kidonakis, G.\ Oderda and G.\ Sterman, \npb{531}{1998}{365}
  [hep-ph/9803241];
  G.\ Oderda \pre{61}{2000}{014004} [hep-ph/9903240]
\bibitem{BCMN} R.\ Bonciani, S.\ Catani, M.\ Mangano and P.\ Nason,
  \plb{575}{2003}{268} [hep-ph/0307035]
\bibitem{ASZ} A.\ Banfi, G.P.\ Salam and G.\ Zanderighi,
  \plb{584}{2004}{298}

\bibitem{QVscale}
% Thrust in E+E-
S.~Catani, L.~Trentadue, G.~Turnock and B.R.~Webber,
\npb{407}{1993}{3}; \\
% C-PARAMETER IN E+ E- ANNIHILATION.
S. Catani and B.R. Webber, \plb{427}{1998}{377}
[hep-ph/9801350];\\ 
%The Immortal Unappreciated broadening paper
Yu.L.~Dokshitzer, A.~Lucenti, G.~Marchesini and G.P.~Salam,
%``On the {QCD} analysis of jet broadening,''
\jhep{01}{1998}{011} [hep-ph/9801324]

\bibitem{DMunder} 
Yu.L.~Dokshitzer and G.~Marchesini, ``{\em Soft gluons at large angles
in hadron collisions}'' [hep-ph/0509078]

\end{thebibliography}
\end{document}